\documentstyle[epic,preprint,aps,eepic,tighten]{revtex}

\newfont{\ssf}{cmss10 scaled 1000}
\def\bls{\baselineskip12.5pt}

\def\al{\alpha}
\def\at{\tilde{\alpha}}
\def\bet{\beta}
\def\gam{\gamma}
\def\Gam{\Gamma}
\def\dd{\Delta}
\def\ub{\bar{u}}
\def\d0{\Delta_0}
\def\dn{\Delta_0}
\def\dnt{\tilde{\Delta}_0}
\def\D{\omega_D}
\def\s{\sigma}
\def\bs{\hat{\sigma}}
\def\om{\omega}
\def\Om{\Omega}
\def\eps{\epsilon}
\def\epst{\tilde{\epsilon}}
\def\gamt{\tilde{\gamma}}
\def\Gamt{\tilde{\Gamma}}
\def\la{\lambda}
\def\EW{{\langle\sigma_x\rangle}}
\def\EV{{\langle\sigma_z\rangle}}

\def\A{{\cal A}}
\def\B{B}
\def\PP{{\cal PLP}}
\def\PQ{{\cal PLQ}}
\def\QP{{\cal QLP}}
\def\QQ{{\cal QLQ}}
\def\L{{\cal L}}
\def\P{{\cal P}}
\def\Q{{\cal Q}}
\def\H{{\cal H}}
\def\C{{\cal C}}
\def\BC{\hat{C}}
\def\Cr{\stackrel{>}{\cal C}}
\def\Ck{\stackrel{<}{\cal C}}
\def\CT{{C}_{zz}}
\def\CL{{C}_{xx}}
\def\N{{\cal N}}
\def\J{J}
\def\I{{\cal I}}
\def\V{{\cal V}}
\def\F{{\cal F}}
\def\M{{\cal M}}
\def\K{{\cal K}}
\def\dw{\tilde{\Delta}_0(\omega)}
\def\dt{\tilde{\Delta}_0}
\def\wt{\tilde{\omega}}
\def\wc{\omega_c}
\def\omt{\tilde{\omega}}
\def\si{\Sigma}
\def\<{\stackrel{<}{\sim}}
\def\>{\stackrel{>}{\sim}}
\def\gr{ g_+}
\def\gk{ g_-}

\def\d{\mbox{d}}
\def\e{\mbox{e}}
\def\tr{\mbox{tr}}

\def\ba{\begin{eqnarray}}
\def\ea{\end{eqnarray}}
\def\nn{\nonumber}

\begin{document}

\thispagestyle{empty}
\title{RELAXATION DUE TO INCOHERENT TUNNELLING IN
              DIELECTRIC GLASSES}

\author{{\sl Peter Neu}$\quad$ and $\quad${\sl Alois W\"urger}\\[0.4cm]{\it
Institut f\"ur}\\
{\it Theoretische Physik}\\
{\it Universit\"at Heidelberg}\\
{\it Philosophenweg 19}\\ {\it 69120 Heidelberg,  Germany}}

\maketitle

\begin{abstract}
 A new relaxation mechanism is shown to arise from overdamped two-level systems
above a critical temperature $T^*\approx 5$ K, thus yielding an explanation for
experimental observations in dielectric glasses in the temperature range
between $T^*$ and the relaxation peak at 50 K. Using the distribution function
of the tunnelling model for the parameters of the two-level systems, both the
linear
 decrease of the sound velocity and the linear increase of the absorption
 up to the relaxation maximum, are quantitatively accounted for by our theory.
\end{abstract}

$\\$

\noindent PACS. 61.40-- amorphous and polymeric materials $\\$
PACS. 63.50 -- disordered solids, vibrational states $\\$
PACS. 77.22G -- relaxation phenomena, dielectrics

%\newpage

\newpage
Low temperature properties of glasses below 1 Kelvin \cite{HA}
are satisfactorily explained
by the assumption of localised tunnelling states (TS) with a wide distribution
of
energies and relaxation times; these TS are commonly described by
a mapping  on two-level systems (TLS) \cite{AHV}.
 Usually only the direct (one-phonon)
 relaxation mechanism of TS  with phonons is considered \cite{Jae}. For this
reason  there is poor agreement with experiment at
 temperatures above a few Kelvin  \cite{Krau,Ant,Do}; especially the linear
increase
 of the absorption up to the relaxation peak at about 50 K and the
  linear decrease with temperature of the sound
velocity  seem to be a universal characteristic of amorphous substances
\cite{Krau,Ant,Do}.  There are attempts to explain that linear temperature
variation by  thermally activated processes
\cite{merz,KKI}, or by elastic anharmonicity of the lattice \cite{Do}, or by a
modification of the standard distribution function for the
tunnelling parameters \cite{Ant}.

In this letter  we provide an explanation for the temperature
variation above a few Kelvin which relies on the  two-level description and the
standard distribution function, thus avoiding introduction of new parameters.
We find in this temperature regime incoherent tunnelling rather than
coherent oscillations; as a result, all TLS, even the symmetric ones,
contribute to relaxation  and therefore yield  a more pronounced
temperature dependence of sound and microwave propagation.
 In this paper
we extend our previous treatment of symmetric TS \cite{NW} to the biased case.

The model is described by the spin-boson Hamiltonian \cite{Legg}
\ba
H\ = \
   -\,\frac{1}{2}\,\hbar\dn\,\s_x \;+\; \frac{1}{2}\,\hbar\dd\,\s_z\;
     + \;\gam\,e\,\s_z  \;+
	 \; H_B\  ,
\label{sphoas}
\ea
where $\dn$ denotes the tunnelling amplitude, $\dd$ the bias, $\gam$ the
deformation potential,  and
\ba
e \ = \  i\sum_{k}\,k\,\sqrt{\frac{\hbar}{2m_k \om_k}}\; (b_k - b_k^{\dag})
\ea
 the distortion  of the lattice.  We consider the coupling of the TS to
three-dimensional acoustic
phonons  ($\om_k = k v$) described by $H_B =  \sum_k\hbar\om_k b_k^{\dag} b_k$
 and  bosonic operators fulfilling   $[b_k,b_{k'}^{\dag}] = \delta_{k,k'}$. The
model is
 specified by the spectral density  which in Debye approximation
 is given by
 \ba
\J(\om)\ =\  \frac{4}{\hbar}\,\sum_{k}\,\frac{\gam^2\om_k}{2m_k v^2}\,
\delta (\om-\om_k)
\ =\  \gamt^2\;\om^3\; \exp(-\om/\om_D)\ ,
\label{spdichas}
\ea
where $\gamt^2 = \gam^2/(\pi^2\varrho v^5\hbar)$, and $\om_D$ is the
Debye frequency,   $\varrho$ the mass density and $v$ the sound
 velocity. In the tunnelling model the parameters $\dn$ and $\dd$ are
assumed to be distributed according to $P(\dn,\dd)=\bar{P}/\dn$, which is
equivalent to
\ba
P(\eps,r)\,\d\eps\,\d r\ = \ \frac{\bar{P}}{2r\sqrt{1-r}}\, \d\eps\,\d r
\label{DF}
\ea
with a constant $\bar{P}$ and new parameters $r = \dn^2/\eps^2$ and  $\eps =
\sqrt{\dn^2 \,+\, \dd^2}$, where $r_{\rm min} \le r\le 1$.

All dynamical information is contained in the symmetrized two-time
correlation function $C_{zz}(t)$, which is calculated
 in the framework of the Mori-Zwanzig projection
formalism \cite{Mori} using a mode-coupling approximation
\cite{NW,Beck}.
With the projector $\P=\sum_{\al} |\s_{\al})(\s_{\al}|
=\I - \Q$ and the scalar product
$(A|B) = \tr [\rho_{eq} (1/2) (AB + BA)]$, the equilibrium density matrix
$\rho_{eq} = \exp(-\beta H)/\tr(\exp(-\beta H))$, the resolvent matrix
$C_{\al\beta}(z) =
(\s_{\al}|[\L-z]^{-1}|\s_{\beta})$
($\al = x,y,z$)
of the Liouvillian $\L\ast = [H,\ast]/\hbar$ can be written as
\ba
\bigm[z\delta_{\al\bet} - \Om_{\al\bet} + M_{\al\bet}(z)\bigm]
\  C_{\bet\gam}(z)
\ =\ -\ \delta_{\al\gam}\quad,
\label{Mo22}
\ea
 Here $\Om_{\al\bet}=(\s_{\al}|\L|\s_{\bet})$ is the frequency matrix and
 \ba
M_{\al\bet}(z)=\left(\Q\L\,\s_{\al}|\,[\QQ-z]^{-1}\,|\Q\L\,\s_{\bet}\right)
= \left( \begin{array}{ccc} \si_{yy}(z) & -\si_{yx}(z) & 0 \\
 -\si_{xy}(z) & \si_{xx}(z) & 0 \\ 0 & 0 & 0\end{array} \right)
 \label{Mo23}
\ea
 is the damping matrix with the spin-phonon resolvent
$\si_{\al\bet}(z) = \left(e\s_{\al}|\,[\QQ-z]^{-1}\,|e\s_{\bet}\right)$.
In mode-coupling approximation the memory functions are decoupled according to
\cite{NW,Beck}
\ba
\Sigma_\al''(\om) \ = \ C_\al''(\om)\,\ast\, \tilde{J}(\om)\ ,
\label{mod1}
\ea
for $\al = x,$ $y,$ $z, $ $a,$ $s$, where we have defined the bath spectral
function
$\tilde{J}(\om)=J(\om)$ $\,\coth(\beta\hbar\om/2)$, the weighted convolution
integral
\ba
g(\om)\,\ast\,h(\om) \ =\ \int
\frac{\d\om'\,\cosh(\beta\hbar\om/2)}{2\,\cosh(\beta\hbar\om'/2)\,\cosh(\beta\hbar (\om - \om')/2)}
\,g(\om - \om')\,h(\om')
\ea
and the resolvent  functions $C_\al(z) = C_{\al\al}$ for $\al = x,y,z$,  and
$C_a(z)  =  -i\, ( C_{xy}(z) - C_{yx}(z))$ and
$C_s(z)  =   C_{xy}(z) + C_{yx}(z)$.
Here the imaginary parts of the resolvent functions, i. e. the spectral
functions,
 are indicated by a double prime;
the real parts are obtained from these via a Kramers-Kronig relation.
By noting $C_y''(\om) \ = \ (\om\,/\,\dn)^2\, C_z''(\om)$ eq. (\ref{Mo22} -
\ref{mod1}) get closed and can be solved numerically by
iteration. They show a transition from coherent tunnelling, where $C_z''(\om)$
has resonances at $\om\approx\pm\eps$ and $\om = 0$, to incoherent tunnelling
motion,
where the three resonances have merged in one single resonance at $\om = 0$
whose
width narrows with further rising temperature.
In both asymptotic
regimes an analytic solution of eq. (\ref{Mo22} - \ref{mod1})  is possible.

$(i)$ In the  {\it coherent} or weak-coupling regime first Born-approximation
is reliable;
after replacing in (\ref{mod1}) $C_\al''(\om)$ by the free spin-spectral
function,
and also discarding the spin-phonon interaction in $\rho_{eq}$, one easily
derives
the well-known results  (cf. \cite{PiGo})
\ba
 C_z^{\prime\prime}(\om) \ &=& \ \pi\,\EV^2\,
\delta(\om)\ ,\nn\\[0.25cm]
 &+& \frac{1-r}{\cosh^2(\bet\hbar\eps/2)}\,
\frac{\Gam_1}{\om^2 +\Gam_1^2}\ +\
 \frac{r}{2 }\ \left[\
\frac{\Gam_2}{(\om -\eps)^2 + \Gam_2^2} \ +\
\frac{\Gam_2}{(\om +\eps)^2 + \Gam_2^2}\ \right]
\label{aspol1}
\ea
with the usual one-phonon rate
\ba
\Gam_1 \ \equiv \  2\Gam_2\ = \
r\,\frac{\pi}{2}\,\gamt^2\,\eps^3\,\coth(\beta\hbar\eps/2)\ .
\label{coh1}
\ea

$(ii)$ In the {\it incoherent} or strong-coupling regime the full dynamics in
$C_z^{\prime\prime}(\om)$
and $\rho_{eq}$ is kept and the eq. (\ref{Mo22} - \ref{mod1}) are treated
self-consistently.
Off-diagonal correlations like $\si_{xy}(z)$ and spin-polarisations $\EW$,
$\EV$
are now negligible.
  The relevant singularities of $C_z(z)$ are at
$z_\pm = \pm \Om - i\Gam_2$, $z_0 = -i\Gam_1$ where
$\Om \approx \dd $,
$\Gam_2 \approx \Gamt$,
$\Gam_1 \approx \dn^2/\Gamt$
asymptotically.  Here we have identified $\Sigma_{x}(z)$ with   $\Sigma_{y}(z)$
and have  defined $\Gamt$ by
$i\,\Gamt = \Sigma_{x}(z=0)$.
For $\Gamt\gg\eps$ the oscillating poles have zero residue, so that the
asymptotic form
of the relevant spectral function reads
\ba
\ C_z''(\om) \ = \ \frac{\Gam_1}{\om^2\,+\,\Gam_1^2}\ .
\label{qw}
\ea
Inserting this in the mode-coupling integral (\ref{mod1}) yields
\ba
\Gamt\ = \ \frac{\pi^2}{2}\;\gamt\;\left(\frac{k_BT}{\hbar}\right)^2\ .
\label{inc1}
\ea

For low-frequency acoustic experiments on  glasses, only the relaxational pole
at $\om = 0$
is relevant.  For this pole the following  formulae  reasonably  interpolate
 between the behaviour in the coherent  (\ref{aspol1} - \ref{coh1})
and the incoherent regime (\ref{qw} - \ref{inc1})
\ba
C_{\rm rel}^{\prime\prime}(\om)\ = \
\frac{\Gamt^2\,+\,\dd^2/\cosh^2(\beta\hbar\eps/2)}{\Gamt^2\,+\,\eps^2}\;
\frac{\Gam_1}{\om^2\,+\,\Gam_1^2}
\ea
 with the relaxation rate
\ba
			\Gam_1\ = \ \frac{r\,\eps^2\,\Gamt}{\Gamt^2\,+\,\eps^2} \ =:\
\frac{r}{\tau_{\rm min}}\ ,
			\label{g1ipp}
\ea
 $\tau_{\rm min}$  being independent of  $r$, and
\ba
\Gamt\ = \ \left\{\begin{array}{ll}
\frac{\pi}{2}\,\gamt^2\,\eps^3\,\coth(\beta\hbar\eps/2) \ =: \ \Gamt_{\rm 1ph}
& \qquad
T\,<\,T^\ast\\[0.3cm]
\frac{\pi^2}{2}\,\gamt\,\left(\frac{k_BT}{\hbar}\right)^2 \ =: \Gamt_{\rm MC} &
\qquad
T\,\ge\, T^\ast\ .
                              \end{array}\right.
\ea
 Here  $T^\ast$ is the temperature were all thermal TLS ($\hbar\eps \le k_BT$)
 are overdamped. The condition $\tilde{\Gam}_{\rm MC} (T^\ast) \equiv
k_BT^\ast/\hbar$
  yields the transition temperature
 \ba
 T^\ast\ = \ \frac{2\hbar}{\pi^2k_B}\,\frac{1}{\gamt}\ .
 \label{tst}
 \ea

Internal friction  and variation of sound velocity are given by \cite{HA}
\ba
Q^{-1}&=&\frac{\gam^2}{\varrho v^2}\;\overline{\chi''(\om)}\\[0.25cm]
\frac{\delta v}{v} &=& -\,\frac{1}{2}\,\frac{\gam^2}{\varrho
v^2}\;\overline{\chi'(\om)}\ ;
\ea
the absorbative  part  $\chi''(\om)$
of the dynamical susceptibility is related to the fluctuating part of the
spectral function by the fluctuation-dissipation theorem.
   The bar denotes the average over
tunnelling systems with respect to (\ref{DF}).  By cutting the
$\eps$-integration at
$\eps = $max$(k_BT/\hbar,\Gamt)$  one finds for the internal friction
\ba
Q^{-1} \ = \ \left\{\begin{array}{ll}

\frac{8\,\pi\,C\,\gamt^2\,k_B^3\,T^3}{9\,\om\,\hbar^3}
 \qquad &T<T^\ast_{\rm res}\\[0.4cm]
\frac{\pi}{2}\,C\,\Big(\, 1\ -\    \frac{\hbar\Gamt_{\rm MC}}{2 k_BT}
 \, \arctan\left(\frac{2 k_B T}{\hbar\Gamt_{\rm MC}}\right)\,\Big) \\[0.3cm]
 +\  \frac{\pi}{2}\, C\,
\frac{\hbar\Gamt_{\rm MC}}{2 k_BT}\left(\frac{\pi}{2}\;-\;
\arctan\left(\frac{\Gamt_{\rm MC}\,r_{\rm min}}{\om}\right)\,\right)
\label{exp1}\ , \qquad& T>T^\ast_{\rm res}
 \end{array}\right.
\ea
and -- by adding the contribution of the resonant part $\delta v/v|_{\rm res} =
C\ln(T/T_0)$  --
for the change of the sound
velocity
\ba
\frac{\delta v}{v} \ = \ \left\{\begin{array}{ll}
  C\ln\frac{T}{T_0}\qquad &T<T^\ast_{\rm res}\\[0.4cm]
     - \,\frac{1}{2}\,C\, \ln\frac{T}{T_0}\qquad & T^\ast_{\rm
res}<T<T^\ast\\[0.4cm]
   - \, (\frac{1}{4}+\frac{\pi}{16})\,C\, \frac{T}{T^*}\,
\ln\left(\frac{ T^2}{T^*\hbar\om/k_B}\right)\ . \qquad &
 T>T^\ast
                              \end{array}\right.
\ea
with $C = \bar{P}\gam^2/\varrho v^2$. $T^\ast_{\rm res}$ separates the regimes
where the resonant $(T<T^\ast_{\rm res})$ and the relaxational  process
 $(T>T^\ast_{\rm res})$ prevails.
 Below $T^\ast$ one finds the well-known
logarithmic temperature dependence of the sound velocity and the constant
internal friction. At $T = T^\ast$ the temperature dependence
   changes to  a  linear increase in the absorption and a linear decrease in
the sound velocity.

{}From a recent experiment on Suprasil W \cite{Hannes} one finds
  a transition from the plateau to
the linear increase at about $T^\ast\approx  6$ K which corresponds
according to (\ref{tst}) to a deformation potential of about $\gam \approx 2$
eV.    Here we have used $C = 2.8\,\times\,10^{-4}$
 \cite{Hannes}, $\varrho =$ 2.2 g/$\mbox{cm}^3$,  $v_\ell= 5.8\,\times\,
		   10^5$ cm/sec, $v_t = 3.75\,\times\,10^5$ cm/sec
		    \cite{Jae} and $\gam_\ell^2 \approx 2\gam_t^2$ with $\gam^2/v^5 =
			\gam_\ell^2/v_\ell^5 + 2\gam_t^2/v^5$.
With these values we calculate the slope with respect to temperature;
for a comparison with experiment see Table 1. Both the prefactor and the
logarithmic
variation of the sound velocity with frequency show full agreement.

In Fig. 1-2 we have plotted our theoretical results together with the
experimental data for Suprasil W  \cite{Hannes}.
At temperatures between 100 mK and  15 K there is full agreement between
experiment and theory; for both absorption and sound velocity the measured data
could be reproduced
with the same numerical values for $\gamt$ and $C$.  A similarly  good
agreement has been found at other frequencies and in recent
experiments on GeO$_2$ \cite{Sonja}.

Finally, we comment on the absorption peak at about 30 K. According to (19)
we find in our theory $T_{\rm max}\propto\sqrt{\om}$.
However, experimentally the  frequency dependence  of the relaxation maximum
is found to be much weaker  \cite{Hannes}. This would indicate the onset of
thermally activated processes to occur at some temperature below $T_{\rm max}$
which  yields a  logarithmical variation with frequency, rather than the square
 root dependence. For that case the quoted value of $r_{\rm min}$ in Fig. 1 has
no
 physical  relevance. We stress that this does not affect the temperature
variation below $T_{\rm max}$.

In summary, we have shown that the experimental data in the absorption
and the sound velocity up to the relaxation peak  can be explained
in the framework of the tunnelling model. The novel features arise from the
incoherent dynamics of the tunnelling motion at temperatures above $T^*$. In
particular, this provides a new relaxation mechanism which accounts well for
the experimental findings in glasses above 5 K.

\section*{Acknowledgement}
We are grateful to Johannes Classen, Christian Enss and Sonja Rau for
helpful discussions and for kindly communicating experimental data prior to
publication.

\newpage
%\section*{References}
%\addcontentsline{toc}{section}{Literatur}

\newpage

%\vspace{1cm}
\begin{figure}[hbt]
\begin{minipage}[b]{15cm}
 % GNUPLOT: LaTeX picture using EEPIC macros
\setlength{\unitlength}{0.270900pt}
\begin{picture}(1500,900)(0,0)
\tenrm
\thicklines \path(220,219)(240,219)
\thicklines \path(1436,219)(1416,219)
\put(198,219){\makebox(0,0)[r]{2}}
\thicklines \path(220,329)(240,329)
\thicklines \path(1436,329)(1416,329)
\put(198,329){\makebox(0,0)[r]{4}}
\thicklines \path(220,439)(240,439)
\thicklines \path(1436,439)(1416,439)
\put(198,439){\makebox(0,0)[r]{6}}
\thicklines \path(220,548)(240,548)
\thicklines \path(1436,548)(1416,548)
\put(198,548){\makebox(0,0)[r]{8}}
\thicklines \path(220,658)(240,658)
\thicklines \path(1436,658)(1416,658)
\put(198,658){\makebox(0,0)[r]{10}}
\thicklines \path(220,767)(240,767)
\thicklines \path(1436,767)(1416,767)
\put(198,767){\makebox(0,0)[r]{12}}
\thicklines \path(220,877)(240,877)
\thicklines \path(1436,877)(1416,877)
\put(198,877){\makebox(0,0)[r]{14}}
\thicklines \path(220,113)(220,133)
\thicklines \path(220,877)(220,857)
\put(220,68){\makebox(0,0){0.01}}
\thicklines \path(312,113)(312,123)
\thicklines \path(312,877)(312,867)
\put(312,68){\makebox(0,0){}}
\thicklines \path(365,113)(365,123)
\thicklines \path(365,877)(365,867)
\put(365,68){\makebox(0,0){}}
\thicklines \path(403,113)(403,123)
\thicklines \path(403,877)(403,867)
\put(403,68){\makebox(0,0){}}
\thicklines \path(432,113)(432,123)
\thicklines \path(432,877)(432,867)
\put(432,68){\makebox(0,0){}}
\thicklines \path(457,113)(457,123)
\thicklines \path(457,877)(457,867)
\put(457,68){\makebox(0,0){}}
\thicklines \path(477,113)(477,123)
\thicklines \path(477,877)(477,867)
\put(477,68){\makebox(0,0){}}
\thicklines \path(495,113)(495,123)
\thicklines \path(495,877)(495,867)
\put(495,68){\makebox(0,0){}}
\thicklines \path(510,113)(510,123)
\thicklines \path(510,877)(510,867)
\put(510,68){\makebox(0,0){}}
\thicklines \path(524,113)(524,133)
\thicklines \path(524,877)(524,857)
\put(524,68){\makebox(0,0){0.1}}
\thicklines \path(616,113)(616,123)
\thicklines \path(616,877)(616,867)
\put(616,68){\makebox(0,0){}}
\thicklines \path(669,113)(669,123)
\thicklines \path(669,877)(669,867)
\put(669,68){\makebox(0,0){}}
\thicklines \path(707,113)(707,123)
\thicklines \path(707,877)(707,867)
\put(707,68){\makebox(0,0){}}
\thicklines \path(736,113)(736,123)
\thicklines \path(736,877)(736,867)
\put(736,68){\makebox(0,0){}}
\thicklines \path(761,113)(761,123)
\thicklines \path(761,877)(761,867)
\put(761,68){\makebox(0,0){}}
\thicklines \path(781,113)(781,123)
\thicklines \path(781,877)(781,867)
\put(781,68){\makebox(0,0){}}
\thicklines \path(799,113)(799,123)
\thicklines \path(799,877)(799,867)
\put(799,68){\makebox(0,0){}}
\thicklines \path(814,113)(814,123)
\thicklines \path(814,877)(814,867)
\put(814,68){\makebox(0,0){}}
\thicklines \path(828,113)(828,133)
\thicklines \path(828,877)(828,857)
\put(828,68){\makebox(0,0){1}}
\thicklines \path(920,113)(920,123)
\thicklines \path(920,877)(920,867)
\put(920,68){\makebox(0,0){}}
\thicklines \path(973,113)(973,123)
\thicklines \path(973,877)(973,867)
\put(973,68){\makebox(0,0){}}
\thicklines \path(1011,113)(1011,123)
\thicklines \path(1011,877)(1011,867)
\put(1011,68){\makebox(0,0){}}
\thicklines \path(1040,113)(1040,123)
\thicklines \path(1040,877)(1040,867)
\put(1040,68){\makebox(0,0){}}
\thicklines \path(1065,113)(1065,123)
\thicklines \path(1065,877)(1065,867)
\put(1065,68){\makebox(0,0){}}
\thicklines \path(1085,113)(1085,123)
\thicklines \path(1085,877)(1085,867)
\put(1085,68){\makebox(0,0){}}
\thicklines \path(1103,113)(1103,123)
\thicklines \path(1103,877)(1103,867)
\put(1103,68){\makebox(0,0){}}
\thicklines \path(1118,113)(1118,123)
\thicklines \path(1118,877)(1118,867)
\put(1118,68){\makebox(0,0){}}
\thicklines \path(1132,113)(1132,133)
\thicklines \path(1132,877)(1132,857)
\put(1132,68){\makebox(0,0){10}}
\thicklines \path(1224,113)(1224,123)
\thicklines \path(1224,877)(1224,867)
\put(1224,68){\makebox(0,0){}}
\thicklines \path(1277,113)(1277,123)
\thicklines \path(1277,877)(1277,867)
\put(1277,68){\makebox(0,0){}}
\thicklines \path(1315,113)(1315,123)
\thicklines \path(1315,877)(1315,867)
\put(1315,68){\makebox(0,0){}}
\thicklines \path(1344,113)(1344,123)
\thicklines \path(1344,877)(1344,867)
\put(1344,68){\makebox(0,0){}}
\thicklines \path(1369,113)(1369,123)
\thicklines \path(1369,877)(1369,867)
\put(1369,68){\makebox(0,0){}}
\thicklines \path(1389,113)(1389,123)
\thicklines \path(1389,877)(1389,867)
\put(1389,68){\makebox(0,0){}}
\thicklines \path(1407,113)(1407,123)
\thicklines \path(1407,877)(1407,867)
\put(1407,68){\makebox(0,0){}}
\thicklines \path(1422,113)(1422,123)
\thicklines \path(1422,877)(1422,867)
\put(1422,68){\makebox(0,0){}}
\thicklines \path(1436,113)(1436,133)
\thicklines \path(1436,877)(1436,857)
\put(1436,68){\makebox(0,0){100}}
\thicklines \path(220,113)(1436,113)(1436,877)(220,877)(220,113)
\put(-45,495){\makebox(0,0)[l]{\shortstack{$10^4\times Q^{-1}$}}}
\put(828,23){\makebox(0,0){Temperature (K)}}
\put(312,822){\makebox(0,0)[l]{$\tilde{\gamma} = 2.85 \times 10^{-13}$ s}}
\put(312,767){\makebox(0,0)[l]{$f = 11.4$ kHz}}
\put(312,713){\makebox(0,0)[l]{$r_{\rm min} = 1.55\times 10^{-8}$}}
\put(1306,812){\makebox(0,0)[r]{Suprasil W}}
\put(1350,812){\makebox(0,0){$\diamond$}}
\put(221,122){\makebox(0,0){$\diamond$}}
\put(221,122){\makebox(0,0){$\diamond$}}
\put(228,122){\makebox(0,0){$\diamond$}}
\put(231,122){\makebox(0,0){$\diamond$}}
\put(231,123){\makebox(0,0){$\diamond$}}
\put(240,123){\makebox(0,0){$\diamond$}}
\put(245,124){\makebox(0,0){$\diamond$}}
\put(249,125){\makebox(0,0){$\diamond$}}
\put(253,125){\makebox(0,0){$\diamond$}}
\put(258,126){\makebox(0,0){$\diamond$}}
\put(262,126){\makebox(0,0){$\diamond$}}
\put(264,126){\makebox(0,0){$\diamond$}}
\put(266,127){\makebox(0,0){$\diamond$}}
\put(272,127){\makebox(0,0){$\diamond$}}
\put(277,128){\makebox(0,0){$\diamond$}}
\put(282,130){\makebox(0,0){$\diamond$}}
\put(288,130){\makebox(0,0){$\diamond$}}
\put(292,131){\makebox(0,0){$\diamond$}}
\put(296,132){\makebox(0,0){$\diamond$}}
\put(301,133){\makebox(0,0){$\diamond$}}
\put(306,134){\makebox(0,0){$\diamond$}}
\put(312,137){\makebox(0,0){$\diamond$}}
\put(317,138){\makebox(0,0){$\diamond$}}
\put(323,138){\makebox(0,0){$\diamond$}}
\put(325,140){\makebox(0,0){$\diamond$}}
\put(330,143){\makebox(0,0){$\diamond$}}
\put(336,145){\makebox(0,0){$\diamond$}}
\put(344,147){\makebox(0,0){$\diamond$}}
\put(388,162){\makebox(0,0){$\diamond$}}
\put(396,165){\makebox(0,0){$\diamond$}}
\put(403,167){\makebox(0,0){$\diamond$}}
\put(412,171){\makebox(0,0){$\diamond$}}
\put(419,177){\makebox(0,0){$\diamond$}}
\put(428,180){\makebox(0,0){$\diamond$}}
\put(434,184){\makebox(0,0){$\diamond$}}
\put(434,184){\makebox(0,0){$\diamond$}}
\put(436,185){\makebox(0,0){$\diamond$}}
\put(439,188){\makebox(0,0){$\diamond$}}
\put(440,185){\makebox(0,0){$\diamond$}}
\put(442,189){\makebox(0,0){$\diamond$}}
\put(446,192){\makebox(0,0){$\diamond$}}
\put(447,189){\makebox(0,0){$\diamond$}}
\put(450,195){\makebox(0,0){$\diamond$}}
\put(455,198){\makebox(0,0){$\diamond$}}
\put(459,202){\makebox(0,0){$\diamond$}}
\put(463,203){\makebox(0,0){$\diamond$}}
\put(468,209){\makebox(0,0){$\diamond$}}
\put(474,211){\makebox(0,0){$\diamond$}}
\put(479,216){\makebox(0,0){$\diamond$}}
\put(485,221){\makebox(0,0){$\diamond$}}
\put(491,225){\makebox(0,0){$\diamond$}}
\put(495,230){\makebox(0,0){$\diamond$}}
\put(501,234){\makebox(0,0){$\diamond$}}
\put(507,239){\makebox(0,0){$\diamond$}}
\put(513,243){\makebox(0,0){$\diamond$}}
\put(520,250){\makebox(0,0){$\diamond$}}
\put(527,255){\makebox(0,0){$\diamond$}}
\put(535,260){\makebox(0,0){$\diamond$}}
\put(541,266){\makebox(0,0){$\diamond$}}
\put(548,272){\makebox(0,0){$\diamond$}}
\put(554,277){\makebox(0,0){$\diamond$}}
\put(559,280){\makebox(0,0){$\diamond$}}
\put(563,284){\makebox(0,0){$\diamond$}}
\put(569,286){\makebox(0,0){$\diamond$}}
\put(575,291){\makebox(0,0){$\diamond$}}
\put(581,294){\makebox(0,0){$\diamond$}}
\put(588,299){\makebox(0,0){$\diamond$}}
\put(595,304){\makebox(0,0){$\diamond$}}
\put(601,309){\makebox(0,0){$\diamond$}}
\put(607,313){\makebox(0,0){$\diamond$}}
\put(614,315){\makebox(0,0){$\diamond$}}
\put(622,320){\makebox(0,0){$\diamond$}}
\put(632,326){\makebox(0,0){$\diamond$}}
\put(636,325){\makebox(0,0){$\diamond$}}
\put(642,330){\makebox(0,0){$\diamond$}}
\put(648,333){\makebox(0,0){$\diamond$}}
\put(654,336){\makebox(0,0){$\diamond$}}
\put(662,337){\makebox(0,0){$\diamond$}}
\put(669,339){\makebox(0,0){$\diamond$}}
\put(675,342){\makebox(0,0){$\diamond$}}
\put(679,341){\makebox(0,0){$\diamond$}}
\put(685,342){\makebox(0,0){$\diamond$}}
\put(692,344){\makebox(0,0){$\diamond$}}
\put(699,345){\makebox(0,0){$\diamond$}}
\put(701,348){\makebox(0,0){$\diamond$}}
\put(709,349){\makebox(0,0){$\diamond$}}
\put(710,348){\makebox(0,0){$\diamond$}}
\put(711,347){\makebox(0,0){$\diamond$}}
\put(717,348){\makebox(0,0){$\diamond$}}
\put(720,351){\makebox(0,0){$\diamond$}}
\put(726,350){\makebox(0,0){$\diamond$}}
\put(727,351){\makebox(0,0){$\diamond$}}
\put(732,351){\makebox(0,0){$\diamond$}}
\put(735,351){\makebox(0,0){$\diamond$}}
\put(739,351){\makebox(0,0){$\diamond$}}
\put(743,352){\makebox(0,0){$\diamond$}}
\put(746,352){\makebox(0,0){$\diamond$}}
\put(751,353){\makebox(0,0){$\diamond$}}
\put(753,352){\makebox(0,0){$\diamond$}}
\put(760,352){\makebox(0,0){$\diamond$}}
\put(761,352){\makebox(0,0){$\diamond$}}
\put(769,352){\makebox(0,0){$\diamond$}}
\put(770,350){\makebox(0,0){$\diamond$}}
\put(777,351){\makebox(0,0){$\diamond$}}
\put(780,353){\makebox(0,0){$\diamond$}}
\put(786,354){\makebox(0,0){$\diamond$}}
\put(789,355){\makebox(0,0){$\diamond$}}
\put(795,353){\makebox(0,0){$\diamond$}}
\put(799,353){\makebox(0,0){$\diamond$}}
\put(803,353){\makebox(0,0){$\diamond$}}
\put(807,354){\makebox(0,0){$\diamond$}}
\put(811,352){\makebox(0,0){$\diamond$}}
\put(814,351){\makebox(0,0){$\diamond$}}
\put(819,354){\makebox(0,0){$\diamond$}}
\put(820,357){\makebox(0,0){$\diamond$}}
\put(825,354){\makebox(0,0){$\diamond$}}
\put(826,354){\makebox(0,0){$\diamond$}}
\put(831,354){\makebox(0,0){$\diamond$}}
\put(838,355){\makebox(0,0){$\diamond$}}
\put(848,356){\makebox(0,0){$\diamond$}}
\put(852,356){\makebox(0,0){$\diamond$}}
\put(858,358){\makebox(0,0){$\diamond$}}
\put(863,356){\makebox(0,0){$\diamond$}}
\put(870,358){\makebox(0,0){$\diamond$}}
\put(875,358){\makebox(0,0){$\diamond$}}
\put(882,358){\makebox(0,0){$\diamond$}}
\put(886,358){\makebox(0,0){$\diamond$}}
\put(895,358){\makebox(0,0){$\diamond$}}
\put(899,358){\makebox(0,0){$\diamond$}}
\put(902,359){\makebox(0,0){$\diamond$}}
\put(902,357){\makebox(0,0){$\diamond$}}
\put(906,357){\makebox(0,0){$\diamond$}}
\put(913,357){\makebox(0,0){$\diamond$}}
\put(916,357){\makebox(0,0){$\diamond$}}
\put(922,357){\makebox(0,0){$\diamond$}}
\put(927,355){\makebox(0,0){$\diamond$}}
\put(932,355){\makebox(0,0){$\diamond$}}
\put(937,355){\makebox(0,0){$\diamond$}}
\put(945,356){\makebox(0,0){$\diamond$}}
\put(945,354){\makebox(0,0){$\diamond$}}
\put(954,354){\makebox(0,0){$\diamond$}}
\put(954,355){\makebox(0,0){$\diamond$}}
\put(973,355){\makebox(0,0){$\diamond$}}
\put(973,355){\makebox(0,0){$\diamond$}}
\put(982,356){\makebox(0,0){$\diamond$}}
\put(982,355){\makebox(0,0){$\diamond$}}
\put(990,357){\makebox(0,0){$\diamond$}}
\put(990,358){\makebox(0,0){$\diamond$}}
\put(997,359){\makebox(0,0){$\diamond$}}
\put(997,359){\makebox(0,0){$\diamond$}}
\put(1004,362){\makebox(0,0){$\diamond$}}
\put(1013,366){\makebox(0,0){$\diamond$}}
\put(1018,376){\makebox(0,0){$\diamond$}}
\put(1019,376){\makebox(0,0){$\diamond$}}
\put(1020,368){\makebox(0,0){$\diamond$}}
\put(1022,377){\makebox(0,0){$\diamond$}}
\put(1025,377){\makebox(0,0){$\diamond$}}
\put(1027,372){\makebox(0,0){$\diamond$}}
\put(1029,378){\makebox(0,0){$\diamond$}}
\put(1032,379){\makebox(0,0){$\diamond$}}
\put(1035,381){\makebox(0,0){$\diamond$}}
\put(1040,382){\makebox(0,0){$\diamond$}}
\put(1044,387){\makebox(0,0){$\diamond$}}
\put(1048,391){\makebox(0,0){$\diamond$}}
\put(1054,396){\makebox(0,0){$\diamond$}}
\put(1057,400){\makebox(0,0){$\diamond$}}
\put(1062,406){\makebox(0,0){$\diamond$}}
\put(1067,410){\makebox(0,0){$\diamond$}}
\put(1071,415){\makebox(0,0){$\diamond$}}
\put(1077,420){\makebox(0,0){$\diamond$}}
\put(1082,428){\makebox(0,0){$\diamond$}}
\put(1087,438){\makebox(0,0){$\diamond$}}
\put(1092,449){\makebox(0,0){$\diamond$}}
\put(1130,510){\makebox(0,0){$\diamond$}}
\put(1135,514){\makebox(0,0){$\diamond$}}
\put(1138,518){\makebox(0,0){$\diamond$}}
\put(1144,533){\makebox(0,0){$\diamond$}}
\put(1151,542){\makebox(0,0){$\diamond$}}
\put(1156,545){\makebox(0,0){$\diamond$}}
\put(1166,569){\makebox(0,0){$\diamond$}}
\put(1174,577){\makebox(0,0){$\diamond$}}
\put(1182,591){\makebox(0,0){$\diamond$}}
\put(1189,604){\makebox(0,0){$\diamond$}}
\put(1195,619){\makebox(0,0){$\diamond$}}
\put(1201,634){\makebox(0,0){$\diamond$}}
\put(1205,643){\makebox(0,0){$\diamond$}}
\put(1211,659){\makebox(0,0){$\diamond$}}
\put(1217,673){\makebox(0,0){$\diamond$}}
\put(1229,697){\makebox(0,0){$\diamond$}}
\put(1234,699){\makebox(0,0){$\diamond$}}
\put(1240,707){\makebox(0,0){$\diamond$}}
\put(1246,715){\makebox(0,0){$\diamond$}}
\put(1251,721){\makebox(0,0){$\diamond$}}
\put(1257,729){\makebox(0,0){$\diamond$}}
\put(1263,733){\makebox(0,0){$\diamond$}}
\put(1266,730){\makebox(0,0){$\diamond$}}
\put(1271,733){\makebox(0,0){$\diamond$}}
\put(1275,730){\makebox(0,0){$\diamond$}}
\put(1278,725){\makebox(0,0){$\diamond$}}
\put(1281,723){\makebox(0,0){$\diamond$}}
\put(1286,717){\makebox(0,0){$\diamond$}}
\put(1290,704){\makebox(0,0){$\diamond$}}
\put(1294,697){\makebox(0,0){$\diamond$}}
\put(1298,686){\makebox(0,0){$\diamond$}}
\put(1303,671){\makebox(0,0){$\diamond$}}
\put(1308,658){\makebox(0,0){$\diamond$}}
\put(1312,646){\makebox(0,0){$\diamond$}}
\put(1316,636){\makebox(0,0){$\diamond$}}
\put(1316,616){\makebox(0,0){$\diamond$}}
\put(1320,607){\makebox(0,0){$\diamond$}}
\put(1323,594){\makebox(0,0){$\diamond$}}
\put(1326,583){\makebox(0,0){$\diamond$}}
\put(1329,573){\makebox(0,0){$\diamond$}}
\put(1332,562){\makebox(0,0){$\diamond$}}
\put(1334,552){\makebox(0,0){$\diamond$}}
\put(1337,541){\makebox(0,0){$\diamond$}}
\put(1339,533){\makebox(0,0){$\diamond$}}
\put(1342,523){\makebox(0,0){$\diamond$}}
\put(1344,514){\makebox(0,0){$\diamond$}}
\put(1346,505){\makebox(0,0){$\diamond$}}
\put(1348,497){\makebox(0,0){$\diamond$}}
\put(1350,488){\makebox(0,0){$\diamond$}}
\put(1352,480){\makebox(0,0){$\diamond$}}
\put(1354,471){\makebox(0,0){$\diamond$}}
\put(1356,464){\makebox(0,0){$\diamond$}}
\put(1357,456){\makebox(0,0){$\diamond$}}
\put(1359,448){\makebox(0,0){$\diamond$}}
\put(1361,440){\makebox(0,0){$\diamond$}}
\put(1363,434){\makebox(0,0){$\diamond$}}
\put(1365,426){\makebox(0,0){$\diamond$}}
\put(1367,420){\makebox(0,0){$\diamond$}}
\put(1368,413){\makebox(0,0){$\diamond$}}
\put(1370,406){\makebox(0,0){$\diamond$}}
\put(1371,400){\makebox(0,0){$\diamond$}}
\put(1373,393){\makebox(0,0){$\diamond$}}
\put(1374,388){\makebox(0,0){$\diamond$}}
\put(1376,383){\makebox(0,0){$\diamond$}}
\put(1377,376){\makebox(0,0){$\diamond$}}
\put(1378,371){\makebox(0,0){$\diamond$}}
\put(1380,367){\makebox(0,0){$\diamond$}}
\put(1381,362){\makebox(0,0){$\diamond$}}
\put(1383,358){\makebox(0,0){$\diamond$}}
\put(1384,352){\makebox(0,0){$\diamond$}}
\put(1385,348){\makebox(0,0){$\diamond$}}
\put(1387,343){\makebox(0,0){$\diamond$}}
\put(1388,343){\makebox(0,0){$\diamond$}}
\put(1390,334){\makebox(0,0){$\diamond$}}
\put(1391,337){\makebox(0,0){$\diamond$}}
\put(1392,334){\makebox(0,0){$\diamond$}}
\put(1394,328){\makebox(0,0){$\diamond$}}
\put(1395,323){\makebox(0,0){$\diamond$}}
\put(1396,320){\makebox(0,0){$\diamond$}}
\put(1400,317){\makebox(0,0){$\diamond$}}
\put(1402,309){\makebox(0,0){$\diamond$}}
\put(1403,312){\makebox(0,0){$\diamond$}}
\put(1404,287){\makebox(0,0){$\diamond$}}
\put(1405,280){\makebox(0,0){$\diamond$}}
\put(1407,277){\makebox(0,0){$\diamond$}}
\put(1410,265){\makebox(0,0){$\diamond$}}
\put(1412,258){\makebox(0,0){$\diamond$}}
\put(1414,253){\makebox(0,0){$\diamond$}}
\put(1415,254){\makebox(0,0){$\diamond$}}
\put(1415,253){\makebox(0,0){$\diamond$}}
\put(1416,250){\makebox(0,0){$\diamond$}}
\put(1417,245){\makebox(0,0){$\diamond$}}
\put(1418,241){\makebox(0,0){$\diamond$}}
\put(1419,239){\makebox(0,0){$\diamond$}}
\put(1420,236){\makebox(0,0){$\diamond$}}
\put(1421,234){\makebox(0,0){$\diamond$}}
\put(1423,228){\makebox(0,0){$\diamond$}}
\put(1424,225){\makebox(0,0){$\diamond$}}
\put(1424,222){\makebox(0,0){$\diamond$}}
\put(1426,218){\makebox(0,0){$\diamond$}}
\put(1427,216){\makebox(0,0){$\diamond$}}
\put(1427,214){\makebox(0,0){$\diamond$}}
\put(1429,212){\makebox(0,0){$\diamond$}}
\put(1429,209){\makebox(0,0){$\diamond$}}
\put(1430,208){\makebox(0,0){$\diamond$}}
\put(1431,205){\makebox(0,0){$\diamond$}}
\put(1432,199){\makebox(0,0){$\diamond$}}
\put(1432,203){\makebox(0,0){$\diamond$}}
\put(1433,199){\makebox(0,0){$\diamond$}}
\put(1434,199){\makebox(0,0){$\diamond$}}
\put(1434,197){\makebox(0,0){$\diamond$}}
\put(1436,197){\makebox(0,0){$\diamond$}}
\put(1306,767){\makebox(0,0)[r]{theory}}
\thinlines \path(1328,767)(1394,767)
\thinlines
\path(353,113)(355,113)(367,114)(380,116)(392,117)(404,120)(417,123)(429,127)(441,132)(453,138)(466,146)(478,156)(490,167)(503,180)(515,195)(527,211)(539,227)(552,244)(564,260)(576,275)(588,289)(601,301)(613,311)(625,320)(638,326)(650,332)(662,336)(674,340)(687,342)(699,344)(711,346)(724,347)(736,348)(748,349)(760,350)(773,350)(785,351)(797,351)(810,352)(822,352)(834,353)(846,353)(859,354)(871,354)(883,355)(896,356)(908,357)(920,358)(932,360)(945,361)(957,364)
\thinlines
\path(957,364)(969,366)(982,369)(994,373)(1006,377)(1018,382)(1031,388)(1043,395)(1055,403)(1068,412)(1080,423)(1092,436)(1104,450)(1117,466)(1129,484)(1141,504)(1153,526)(1166,550)(1178,575)(1190,602)(1203,628)(1215,654)(1227,678)(1239,699)(1252,715)(1264,725)(1276,728)(1289,722)(1301,708)(1313,687)(1325,660)(1338,628)(1350,594)(1362,559)(1375,524)(1387,491)(1399,459)(1411,430)(1424,402)(1436,377)
\end{picture}
\end{minipage}
\end{figure}

\vspace{2cm}

%\vspace{1cm}
\begin{figure}[hbt]
\begin{minipage}[b]{15cm}
 % GNUPLOT: LaTeX picture using EEPIC macros
\setlength{\unitlength}{0.270900pt}
\begin{picture}(1500,900)(0,0)
\tenrm
\thicklines \path(220,113)(240,113)
\thicklines \path(1436,113)(1416,113)
\put(198,113){\makebox(0,0)[r]{4}}
\thicklines \path(220,209)(240,209)
\thicklines \path(1436,209)(1416,209)
\put(198,209){\makebox(0,0)[r]{5}}
\thicklines \path(220,304)(240,304)
\thicklines \path(1436,304)(1416,304)
\put(198,304){\makebox(0,0)[r]{6}}
\thicklines \path(220,400)(240,400)
\thicklines \path(1436,400)(1416,400)
\put(198,400){\makebox(0,0)[r]{7}}
\thicklines \path(220,495)(240,495)
\thicklines \path(1436,495)(1416,495)
\put(198,495){\makebox(0,0)[r]{8}}
\thicklines \path(220,591)(240,591)
\thicklines \path(1436,591)(1416,591)
\put(198,591){\makebox(0,0)[r]{9}}
\thicklines \path(220,686)(240,686)
\thicklines \path(1436,686)(1416,686)
\put(198,686){\makebox(0,0)[r]{10}}
\thicklines \path(220,782)(240,782)
\thicklines \path(1436,782)(1416,782)
\put(198,782){\makebox(0,0)[r]{11}}
\thicklines \path(220,877)(240,877)
\thicklines \path(1436,877)(1416,877)
\put(198,877){\makebox(0,0)[r]{12}}
\thicklines \path(340,113)(340,133)
\thicklines \path(340,877)(340,857)
\put(340,68){\makebox(0,0){2}}
\thicklines \path(462,113)(462,133)
\thicklines \path(462,877)(462,857)
\put(462,68){\makebox(0,0){4}}
\thicklines \path(584,113)(584,133)
\thicklines \path(584,877)(584,857)
\put(584,68){\makebox(0,0){6}}
\thicklines \path(705,113)(705,133)
\thicklines \path(705,877)(705,857)
\put(705,68){\makebox(0,0){8}}
\thicklines \path(827,113)(827,133)
\thicklines \path(827,877)(827,857)
\put(827,68){\makebox(0,0){10}}
\thicklines \path(949,113)(949,133)
\thicklines \path(949,877)(949,857)
\put(949,68){\makebox(0,0){12}}
\thicklines \path(1071,113)(1071,133)
\thicklines \path(1071,877)(1071,857)
\put(1071,68){\makebox(0,0){14}}
\thicklines \path(1192,113)(1192,133)
\thicklines \path(1192,877)(1192,857)
\put(1192,68){\makebox(0,0){16}}
\thicklines \path(1314,113)(1314,133)
\thicklines \path(1314,877)(1314,857)
\put(1314,68){\makebox(0,0){18}}
\thicklines \path(1436,113)(1436,133)
\thicklines \path(1436,877)(1436,857)
\put(1436,68){\makebox(0,0){20}}
\thicklines \path(220,113)(1436,113)(1436,877)(220,877)(220,113)
\put(-45,495){\makebox(0,0)[l]{\shortstack{$10^3\times\delta v/v$}}}
\put(828,23){\makebox(0,0){Temperature (K)}}
\put(979,686){\makebox(0,0)[l]{$\tilde{\gamma} = 2.85\times 10^{-13}$ s}}
\put(979,591){\makebox(0,0)[l]{$f = 11.4$ kHz}}
\put(1306,812){\makebox(0,0)[r]{Suprasil W}}
\put(1350,812){\makebox(0,0){$\diamond$}}
\put(220,799){\makebox(0,0){$\diamond$}}
\put(220,800){\makebox(0,0){$\diamond$}}
\put(220,801){\makebox(0,0){$\diamond$}}
\put(220,802){\makebox(0,0){$\diamond$}}
\put(220,802){\makebox(0,0){$\diamond$}}
\put(220,803){\makebox(0,0){$\diamond$}}
\put(220,803){\makebox(0,0){$\diamond$}}
\put(221,804){\makebox(0,0){$\diamond$}}
\put(221,805){\makebox(0,0){$\diamond$}}
\put(221,805){\makebox(0,0){$\diamond$}}
\put(221,805){\makebox(0,0){$\diamond$}}
\put(221,806){\makebox(0,0){$\diamond$}}
\put(221,806){\makebox(0,0){$\diamond$}}
\put(221,807){\makebox(0,0){$\diamond$}}
\put(221,807){\makebox(0,0){$\diamond$}}
\put(221,808){\makebox(0,0){$\diamond$}}
\put(221,808){\makebox(0,0){$\diamond$}}
\put(221,808){\makebox(0,0){$\diamond$}}
\put(222,809){\makebox(0,0){$\diamond$}}
\put(222,810){\makebox(0,0){$\diamond$}}
\put(222,810){\makebox(0,0){$\diamond$}}
\put(222,810){\makebox(0,0){$\diamond$}}
\put(222,811){\makebox(0,0){$\diamond$}}
\put(222,811){\makebox(0,0){$\diamond$}}
\put(222,811){\makebox(0,0){$\diamond$}}
\put(222,811){\makebox(0,0){$\diamond$}}
\put(223,811){\makebox(0,0){$\diamond$}}
\put(223,812){\makebox(0,0){$\diamond$}}
\put(224,813){\makebox(0,0){$\diamond$}}
\put(224,813){\makebox(0,0){$\diamond$}}
\put(225,813){\makebox(0,0){$\diamond$}}
\put(225,813){\makebox(0,0){$\diamond$}}
\put(225,813){\makebox(0,0){$\diamond$}}
\put(226,813){\makebox(0,0){$\diamond$}}
\put(226,813){\makebox(0,0){$\diamond$}}
\put(226,813){\makebox(0,0){$\diamond$}}
\put(227,812){\makebox(0,0){$\diamond$}}
\put(227,812){\makebox(0,0){$\diamond$}}
\put(228,812){\makebox(0,0){$\diamond$}}
\put(228,811){\makebox(0,0){$\diamond$}}
\put(229,811){\makebox(0,0){$\diamond$}}
\put(229,811){\makebox(0,0){$\diamond$}}
\put(230,810){\makebox(0,0){$\diamond$}}
\put(230,810){\makebox(0,0){$\diamond$}}
\put(231,809){\makebox(0,0){$\diamond$}}
\put(232,808){\makebox(0,0){$\diamond$}}
\put(232,808){\makebox(0,0){$\diamond$}}
\put(233,807){\makebox(0,0){$\diamond$}}
\put(234,807){\makebox(0,0){$\diamond$}}
\put(234,806){\makebox(0,0){$\diamond$}}
\put(236,805){\makebox(0,0){$\diamond$}}
\put(236,804){\makebox(0,0){$\diamond$}}
\put(237,803){\makebox(0,0){$\diamond$}}
\put(238,803){\makebox(0,0){$\diamond$}}
\put(239,802){\makebox(0,0){$\diamond$}}
\put(240,801){\makebox(0,0){$\diamond$}}
\put(241,800){\makebox(0,0){$\diamond$}}
\put(241,800){\makebox(0,0){$\diamond$}}
\put(243,799){\makebox(0,0){$\diamond$}}
\put(243,799){\makebox(0,0){$\diamond$}}
\put(243,799){\makebox(0,0){$\diamond$}}
\put(245,798){\makebox(0,0){$\diamond$}}
\put(245,797){\makebox(0,0){$\diamond$}}
\put(246,797){\makebox(0,0){$\diamond$}}
\put(247,796){\makebox(0,0){$\diamond$}}
\put(248,796){\makebox(0,0){$\diamond$}}
\put(248,795){\makebox(0,0){$\diamond$}}
\put(249,795){\makebox(0,0){$\diamond$}}
\put(250,794){\makebox(0,0){$\diamond$}}
\put(251,794){\makebox(0,0){$\diamond$}}
\put(252,793){\makebox(0,0){$\diamond$}}
\put(253,793){\makebox(0,0){$\diamond$}}
\put(255,792){\makebox(0,0){$\diamond$}}
\put(255,792){\makebox(0,0){$\diamond$}}
\put(257,791){\makebox(0,0){$\diamond$}}
\put(257,790){\makebox(0,0){$\diamond$}}
\put(260,789){\makebox(0,0){$\diamond$}}
\put(260,788){\makebox(0,0){$\diamond$}}
\put(263,788){\makebox(0,0){$\diamond$}}
\put(264,787){\makebox(0,0){$\diamond$}}
\put(266,787){\makebox(0,0){$\diamond$}}
\put(267,786){\makebox(0,0){$\diamond$}}
\put(268,786){\makebox(0,0){$\diamond$}}
\put(270,784){\makebox(0,0){$\diamond$}}
\put(272,784){\makebox(0,0){$\diamond$}}
\put(273,783){\makebox(0,0){$\diamond$}}
\put(275,782){\makebox(0,0){$\diamond$}}
\put(275,782){\makebox(0,0){$\diamond$}}
\put(278,781){\makebox(0,0){$\diamond$}}
\put(278,780){\makebox(0,0){$\diamond$}}
\put(281,780){\makebox(0,0){$\diamond$}}
\put(284,779){\makebox(0,0){$\diamond$}}
\put(289,777){\makebox(0,0){$\diamond$}}
\put(291,776){\makebox(0,0){$\diamond$}}
\put(294,774){\makebox(0,0){$\diamond$}}
\put(298,773){\makebox(0,0){$\diamond$}}
\put(302,771){\makebox(0,0){$\diamond$}}
\put(305,769){\makebox(0,0){$\diamond$}}
\put(310,767){\makebox(0,0){$\diamond$}}
\put(313,766){\makebox(0,0){$\diamond$}}
\put(319,763){\makebox(0,0){$\diamond$}}
\put(322,762){\makebox(0,0){$\diamond$}}
\put(325,761){\makebox(0,0){$\diamond$}}
\put(325,762){\makebox(0,0){$\diamond$}}
\put(328,760){\makebox(0,0){$\diamond$}}
\put(334,757){\makebox(0,0){$\diamond$}}
\put(337,755){\makebox(0,0){$\diamond$}}
\put(342,752){\makebox(0,0){$\diamond$}}
\put(347,750){\makebox(0,0){$\diamond$}}
\put(352,747){\makebox(0,0){$\diamond$}}
\put(357,745){\makebox(0,0){$\diamond$}}
\put(366,741){\makebox(0,0){$\diamond$}}
\put(366,739){\makebox(0,0){$\diamond$}}
\put(376,734){\makebox(0,0){$\diamond$}}
\put(376,734){\makebox(0,0){$\diamond$}}
\put(401,717){\makebox(0,0){$\diamond$}}
\put(401,716){\makebox(0,0){$\diamond$}}
\put(413,709){\makebox(0,0){$\diamond$}}
\put(413,709){\makebox(0,0){$\diamond$}}
\put(425,703){\makebox(0,0){$\diamond$}}
\put(425,703){\makebox(0,0){$\diamond$}}
\put(437,696){\makebox(0,0){$\diamond$}}
\put(437,695){\makebox(0,0){$\diamond$}}
\put(450,689){\makebox(0,0){$\diamond$}}
\put(466,677){\makebox(0,0){$\diamond$}}
\put(475,674){\makebox(0,0){$\diamond$}}
\put(476,674){\makebox(0,0){$\diamond$}}
\put(479,672){\makebox(0,0){$\diamond$}}
\put(484,670){\makebox(0,0){$\diamond$}}
\put(489,667){\makebox(0,0){$\diamond$}}
\put(494,664){\makebox(0,0){$\diamond$}}
\put(497,663){\makebox(0,0){$\diamond$}}
\put(504,659){\makebox(0,0){$\diamond$}}
\put(511,655){\makebox(0,0){$\diamond$}}
\put(521,650){\makebox(0,0){$\diamond$}}
\put(531,643){\makebox(0,0){$\diamond$}}
\put(541,638){\makebox(0,0){$\diamond$}}
\put(554,629){\makebox(0,0){$\diamond$}}
\put(563,625){\makebox(0,0){$\diamond$}}
\put(578,615){\makebox(0,0){$\diamond$}}
\put(589,608){\makebox(0,0){$\diamond$}}
\put(602,600){\makebox(0,0){$\diamond$}}
\put(620,589){\makebox(0,0){$\diamond$}}
\put(635,580){\makebox(0,0){$\diamond$}}
\put(652,570){\makebox(0,0){$\diamond$}}
\put(668,561){\makebox(0,0){$\diamond$}}
\put(683,553){\makebox(0,0){$\diamond$}}
\put(696,543){\makebox(0,0){$\diamond$}}
\put(717,527){\makebox(0,0){$\diamond$}}
\put(735,515){\makebox(0,0){$\diamond$}}
\put(800,475){\makebox(0,0){$\diamond$}}
\put(819,464){\makebox(0,0){$\diamond$}}
\put(840,452){\makebox(0,0){$\diamond$}}
\put(854,443){\makebox(0,0){$\diamond$}}
\put(886,426){\makebox(0,0){$\diamond$}}
\put(921,405){\makebox(0,0){$\diamond$}}
\put(948,397){\makebox(0,0){$\diamond$}}
\put(1004,353){\makebox(0,0){$\diamond$}}
\put(1058,325){\makebox(0,0){$\diamond$}}
\put(1108,296){\makebox(0,0){$\diamond$}}
\put(1157,266){\makebox(0,0){$\diamond$}}
\put(1196,245){\makebox(0,0){$\diamond$}}
\put(1244,220){\makebox(0,0){$\diamond$}}
\put(1278,201){\makebox(0,0){$\diamond$}}
\put(1329,173){\makebox(0,0){$\diamond$}}
\put(1380,146){\makebox(0,0){$\diamond$}}
\put(1306,767){\makebox(0,0)[r]{theory}}
\thinlines \path(1328,767)(1394,767)
\thinlines
\path(220,800)(220,800)(232,797)(245,792)(257,787)(269,782)(281,777)(294,771)(306,765)(318,759)(331,753)(343,747)(355,741)(367,735)(380,729)(392,722)(404,716)(417,709)(429,703)(441,696)(453,690)(466,683)(478,676)(490,670)(503,663)(515,656)(527,649)(539,642)(552,635)(564,628)(576,621)(588,614)(601,607)(613,600)(625,593)(638,586)(650,579)(662,572)(674,564)(687,557)(699,550)(711,543)(724,535)(736,528)(748,521)(760,513)(773,506)(785,499)(797,491)(810,484)(822,476)
\thinlines
\path(822,476)(834,469)(846,461)(859,454)(871,446)(883,439)(896,431)(908,424)(920,416)(932,409)(945,401)(957,393)(969,386)(982,378)(994,370)(1006,363)(1018,355)(1031,347)(1043,340)(1055,332)(1068,324)(1080,316)(1092,309)(1104,301)(1117,293)(1129,285)(1141,277)(1153,270)(1166,262)(1178,254)(1190,246)(1203,238)(1215,230)(1227,222)(1239,215)(1252,207)(1264,199)(1276,191)(1289,183)(1301,175)(1313,167)(1325,159)(1338,151)(1350,143)(1362,135)(1375,127)(1387,119)(1396,113)
\end{picture}
\end{minipage}
\end{figure}
 \newpage

\section*{Captions}

\noindent Fig. 1:  Internal friction for Suprasil W at 11.4 kHz. The data
($\diamond$)
are from Classen et al. \cite{Hannes}. In our theory (--) we have used $C =
2.8\times 10^{-4}$
and   $\gam_\ell = 2.2$ eV

\noindent Fig. 2: Relative variation of the sound velocity for Suprasil W at
11.4 kHz. The data ($\diamond$)
are from Classen et al. \cite{Hannes}. In our theory (--) we have used the same
numerical value for $C$ and $\gam_\ell$ as in Fig. 1

\begin{table} {Table 1: Comparison of theoretical and experimental  results for
Suprasil W
in the incoherent regime $T>T^\ast$. The experimental values are taken
from \cite{Hannes}}\\[0.5ex]
\ba
\begin{array}{|c||c|c|} \hline
 & \mbox{Experiment}&\mbox{Theory}\\[1ex] \hline
Q^{-1} & (7 \pm 2) \times 10^{-5} T/\mbox{K}
& \quad  6.1 \times 10^{-5} T/\mbox{K}
\quad\nn \\[1ex] \hline
\delta v/v &   2\times  10^{-5} T/\mbox{K}\,\ln(1.6\times
10^{-13}\om/2\pi\mbox{Hz})\nn &
  \quad 2.2\times  10^{-5} T/\mbox{K}\,\ln(T^\ast\hbar\om/(k_BT^2)) \nn
  \quad \\[1ex] \hline
\end{array}
\ea
\end{table}


\begin{thebibliography}{12}
\bibliographystyle{unsrt}
%%%%%%%%%%%%%%%%%%%%%%%%%%%%%%%%%%%%%%%%


\bibitem{HA} Hunklinger, S., Arnold, W.: In: Physical acoustics, vol. 12,
                      Thurston R. N., Mason W. P. (eds). New York: Academic
Press  (1976);
 Hunklinger, S., Raychaudhuri, A. K.: In: Progress in low temperature
                      physics, vol. IX, Brewer D. F. (ed.). Amsterdam: Elsevier
(1986)
\bibitem{AHV} Anderson,  P. W., Halperin, B. I., Varma,  C.: Philos. Mag. {\bf
25}, 1 (1972);
Phillips,  W. A.: J. Low. Temp. Phys. {\bf 7}, 351 (1972)
\bibitem{Jae} J\"ackle,  J.: Z. Phys.  {\bf  257}, 212 (1972)
\bibitem{Krau} Krause, J. T.: J. Appl. Phys. {\bf 42}, 3035 (1971);
Bellessa, G., Lemercier, C., Caldemaison, D.: Phys. Lett. {\bf 62A},
                       127 (1977); Bellesa, G.: Phys. Rev. Lett. {\bf 40}, 1456
(1978)
\bibitem{Ant} Anthony, P. J., Anderson, A. C.: Phys. Rev. {\bf B 20}, 763
(1979)
\bibitem{Do} Doussineau, P., Frenois, C., Leisure, R. G., Levelut, A., Prieur,
J.-Y.:
                     J. Physique {\bf 41}, 1193 (1980)
\bibitem{merz} Tielburger, D., Merz, R., Ehrenfels, R., Hunklinger, S.:
                        Phys. Rev. {\bf B} {\bf 45},
                           2750 (1992)
\bibitem{KKI}  Buchenau, U., Galperin, Yu. M., Gurevich, V. L.,
 Parshin, D. A.,  Ramos, M. A., Schober, H.R.:  Phys. Rev {\bf B} {\bf 46},
                         2798 (1992)
\bibitem{NW} Neu, P., W\"urger, A.:  appears in Z. Phys. {\bf B}
\bibitem{Legg}  Leggett, A. J.,  Chakravarty, S.,  Dorsey, A. T.,
  Fisher, M. P. A., Garg, A.,  Zwerger, W.: Rev. Mod. Phys. {\bf 59},  1
(1987);
 Weiss, U.: Quantum Dissipative Dynamics, Series in Modern Condensed
                         Matter Physics, Vol. 2, World Scientific, Singapore
(1993)
\bibitem{Mori} Mori, H.: Progr. Theor. Phys. {\bf 33}, 127 (1965);
                       Zwanzig,  R.: J. Chem. Phys. {\bf 33}, 1338 (1960)
\bibitem{Beck} Beck, R., G\"otze, W.,  Prelovsek, P.: Phys. Rev {\bf A} {\bf
20},
                         1140 (1979);
 Zwerger, W.:  Z. Phys.  B {\bf 53}, 53 (1983);  ibid {\bf 54},
                            87 (1983)
\bibitem{PiGo}   Pirc, R.,  Gosar, P.: Phys. kondens. Materie {\bf 9}, 377
(1969)
  \bibitem{Hannes} Classen, J., Enss, C., Bechinger, C., Weiss, G., Hunklinger,
            S.: appears in Annalen der Physik (1994)
\bibitem{Sonja} Rau, S.: Private communication (1994)

%%%%%%%%%%%%%%%%%%%%%%%%%%%%%%%%%%%%%%%%%%%%


\end{thebibliography}
\end{document}